\documentclass{aa}
\usepackage{epsfig,times}
\def\mincir{\raise -2.truept\hbox{\rlap{\hbox{$\sim$}}\raise5.truept \hbox{$<$}\ }}
\def\mincireq{\hbox{\raise0.5ex\hbox{$<\lower1.06ex\hbox{$\kern-1.07em{\sim}$}$}}}
\def\magcir{\raise-2.truept\hbox{\rlap{\hbox{$\sim$}}\raise5.truept \hbox{$>$}\ }}
\def\gr{\kern 2pt\hbox{}^\circ{\kern -2pt K}} 

\def\_{\thinspace}

\begin{document}

\title{Intergalactic absorption and blazar $\gamma$-ray spectra}

\author{Massimo Persic\inst{1} 
and 
Alessandro De Angelis\inst{2,} \inst{3}
}

\offprints{M.P.; e-mail: {\tt persic@oats.inaf.it}}

\institute{
INAF and INFN, via G.B.Tiepolo 11, I-34143 Trieste, Italy
        \and
Universit\`a di Udine and INFN, via delle Scienze 208, I-33100 Udine, Italy
        \and
        Instituto Superior T\'ecnico, Lisboa, Portugal
           }
\date{Received ..................; accepted ...................}

\abstract{The distribution of TeV spectral slopes versus redshift for currently known 
TeV blazars (16 sources with  $z$$\leq$0.21, and one with $z$$>$0.25) is essentially a 
scatter plot with hardly any hint of a global trend. We suggest that this is the outcome 
of two combined effects of intergalactic $\gamma\gamma$ absorption, plus an inherent 
feature of the SSC (synchro-self-Compton) process of blazar emission. First, flux dimming 
introduces a bias that favors detection of progressively more flaring sources at higher 
redshifts. According to mainstream SSC models, more flaring source states imply sources 
with flatter TeV slopes. This results in a structured relation between intrinsic TeV slope 
and redshift. The second effect, spectral steepening by intergalactic absorption, affects 
sources progressively with distance and effectively wipes out the intrinsic slope--redshift 
correlation.

}

\maketitle
\markboth{Persic \& De Angelis: IBL absorption and $\gamma$-ray spectra}{}

\section{Introduction}

With the increasing number of very high energy (VHE) $\gamma$-ray detection of blazars, a broad 
correlation between observed slope and redshift was generally expected, in the sense that higher 
redshift sources would tend to show systematically steeper spectra.

The reason for this is absorption of the emitted VHE emission by the intergalactic background 
light (IBL), i.e. the integrated optical/IR emission from the evolving stellar populations in galaxies 
(e.g., Hauser \& Dwek 2001). In fact, pair creation due to the interaction of a hard photon (with energy 
$E$) with a soft photon (with energy $\epsilon$) is expected provided that $\epsilon E$$>$$m_ec^2$. The 
cross section of the $\gamma\gamma$$\rightarrow$$e^\pm$ interaction, $\sigma_{\gamma \gamma}(E,\epsilon)$ 
(Heitler 1960) is maximized when $\epsilon$$\sim$$2(m_ec^2)^2/E$. Then, if $E$$\sim$1 TeV, pair creation 
will be most likely for $\epsilon$$\sim$$0.5\,(E/{\rm TeV})^{-1}$ eV, which corresponds to a 2$\,\mu$m 
($K$-band) photon. The formalism required to evaluate this effect has been largely developed by Stecker 
and collaborators (e.g., Stecker 1971; Stecker et al. 1992 and 2006). 

\begin{figure}
\vspace{6.0cm}
\includegraphics{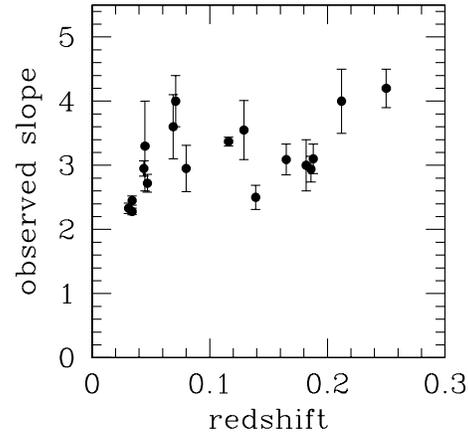}
\caption{Observed TeV slope vs redshift for known TeV blazars. Bars denote 
statistical uncertainties as quoted in the original papers.
}
\end{figure}

However, when we plot the currently known observed TeV slopes against redshift (Fig.1), 
we see a scatter plot with hardly any sign of a global correlation: neglecting slope 
uncertainties, a global linear correlation is suggested only marginally ($\sim$2$\sigma$) 
by the whole data sample ($r$$=$0.52, N=18). Even so, most of the suggestion's strength 
comes from only the two points at $z$$>$0: for the bulk of the distribution, which provides 
a fair sampling of the redshift interval $z$$\leq$0.2, there is no correlation ($r$$=$0.22, 
N=16). In this note we propose an interpretation of Fig.1. 

\begin{table*}
\caption[] {TeV blazar data.}
\begin{flushleft}
\begin{tabular}{ l l l l l l l l l l l }
\hline
\hline
\noalign{\smallskip}
Source  & ~~~~~~$z$  & ~~~~~~~~$\alpha_\gamma$ & ~~~~~~~~F$_\gamma$ & CT & ~~~$\alpha_\gamma^{\rm corr}$ & 
~~~~~~ F$_\gamma^{\rm corr}$ 
&  F$_{5\,{\rm GHz}}$ & $~~~~~~$F$_{\rm x}$ & ~~F$_{x,0}$  & Notes \\
\noalign{\smallskip}
\hline
\noalign{\smallskip}
      (1) & ~~~~ (2)  & ~~~~~~~ (3) & ~~~~~~~~(4) & (5)& ~~~~(6) & ~~~~~~~~(7) & ~~(8) & ~~~~~~(9) & ~~(10) & ~(11) 
\\
\noalign{\smallskip}
\hline
\noalign{\smallskip}
Mrk~421       &  ~~~0.031 & $2.33 \pm 0.08$ & 1.03($\pm$0.03)E-10 & M & ~~~2.15 & ~~~7.59E-11 & 0.730 & ~~~4.0E-10 & 
4.1E-11 & Al+07a \\
Mrk~501       &  ~~~0.034 & $2.28 \pm 0.05$ & 1.71($\pm$0.11)E-11 & M & ~~~2.08 & ~~~1.32E-11 & 1.400 & ~~~4.0E-10 & 3.0E-11 & Al+07b \\
              &           & $2.45 \pm 0.07$ & 3.84($\pm$1.00)E-12 & M & ~~~2.25 & ~~~4.73E-12 &       & ~~~1.7E-10 &         &        \\
PKS~2344+514  &  ~~~0.044 & $2.95 \pm 0.12$ & 1.21($\pm$0.10)E-11 & M & ~~~2.67 & ~~~1.67E-11 & 0.231 & $<$7.0E-11 & 1.7E-11 & Al+07c \\
Mrk~180       &  ~~~0.045 & $3.30 \pm 0.70$ & 8.46($\pm$3.38)E-12 & M & ~~~3.01 & ~~~1.23E-11 & 0.274 & $<$1.0E-10 & 5.0E-12 & Al+06a \\
1ES~1959+650  &  ~~~0.047 & $2.72 \pm 0.14$ & 3.04($\pm$0.35)E-11 & M & ~~~2.41 & ~~~4.63E-11 & 0.253 & ~~~1.4E-10 & 1.4E-11 & Al+06b \\
BL~Lacertae   &  ~~~0.069 & $3.60 \pm 0.50$ & 3.28($\pm$0.26)E-12 & M & ~~~3.12 & ~~~7.15E-12 & 2.490 & ~~~~~~ --- & 5.8E-13 & Al+07d \\
PKS~2005-489  &  ~~~0.071 & $4.00 \pm 0.40$ & 3.32($\pm$0.48)E-12 & H & ~~~3.50 & ~~~7.47E-12 & 1.190 & ~~~4.0E-11 & 0.9E-11 & Ah+05a \\
              &           &                 &                     &   &         &             &       &            &         & P+99 \\
RGB~J0152+017 &  ~~~0.080 & $2.95 \pm 0.36$ & 4.43($\pm$1.24)E-12 & H & ~~~2.38 & ~~~1.15E-11 & 0.050 & ~~~2.7E-12 & 9.1E-13 & Ah+08 \\
              &           &                 &                     &   &         &             &       &            &         & D+01 \\
PKS~2155-304  &  ~~~0.116 & $3.37 \pm 0.07$ & 2.89($\pm$0.18)E-11 & H & ~~~2.51 & ~~~1.36E-10 & 0.310 & ~~~2.7E-11 & 2.0E-11 & Ah+05b \\
1ES~1426+428  &  ~~~0.129 & $3.55 \pm 0.46$ & 2.53($\pm$0.43)E-11 & W & ~~~2.54 & ~~~1.47E-10 & 0.038 & ~~~2.0E-11 & 1.6E-11 & Hor+02 \\
              &           &                 &                     &   &         &             &       &            &         & S+97 \\
1ES~0229+200  &  ~~~0.139 & $2.50 \pm 0.19$ & 4.46($\pm$0.71)E-12 & H & ~~~1.46 & ~~~3.04E-11 & 0.046 & ~~~1.5E-11 & 8.8E-12 & Ah+07c \\
              &           &                 &                     &   &         &             &       &            &         &P+96 \\ 
              &           &                 &                     &   &         &             &       &            &         & D+05 \\
H~2356-309    &  ~~~0.165 & $3.09 \pm 0.24$ & 2.55($\pm$0.68)E-12 & H & ~~~1.84 & ~~~2.66E-11 & 0.065 & ~~~1.0E-11 & 2.4E-11 & Ah+06a \\
              &           &                 &                     &   &         &             &       &            &         & W+84 \\
1ES~1218+304  &  ~~~0.182 & $3.00 \pm 0.40$ & 1.01($\pm$0.26)E-11 & M & ~~~1.61 & ~~~1.39E-10 & 0.056 & ~~~~~~ --- & 1.5E-11 & Al+06c \\
1ES~1101-232  &  ~~~0.186 & $2.94 \pm 0.20$ & 4.35($\pm$0.69)E-12 & H & ~~~1.46 & ~~~6.08E-11 & 0.066 & ~~~5.1E-11 & 2.5E-11 & Ah+07a \\
1ES~0347-121  &  ~~~0.188 & $3.10 \pm 0.23$ & 3.86($\pm$0.73)E-12 & H & ~~~1.67 & ~~~5.77E-11 & 0.008 &~~~2.8E-11  & 6.0E-12 & Ah+07b \\
1ES~1011+496  &  ~~~0.212 & $4.00 \pm 0.50$ & 6.40($\pm$0.32)E-12 & M & ~~~2.37 & ~~~1.43E-09 & 0.636 & ~~~~~~ --- & ~~~ --- & Al+07e \\
PG~1553+113   & $>$0.25   & $4.20 \pm 0.30$ & 5.24($\pm$0.87)E-12 & M & $<$2.28 & $>$2.18E-10 & 0.286 & ~~~5.0E-11 & 1.4E-11 & Al+07f \\
\noalign{\smallskip}
\hline
\hline
\end{tabular}
\end{flushleft}
\smallskip

\noindent
{\it Col.1:} source name.

\noindent
{\it Col.2:} source redshift.

\noindent
{\it Col.3:} observed 0.2-2 TeV photon spectral index, and associated statistical uncertainty. The 
corresponding systematic uncertainties are typically $\sim$0.1 for H.E.S.S. and $\sim$0.2 for MAGIC.

\noindent
{\it Col.4:} observed $>$0.2 TeV flux (in erg cm$^{-2}$ s$^{-1}$), and associated statistical uncertainty 
(from observed spectral normalization only). 

\noindent
{\it Col.5:} Cherenkov telescope (CT) or array with which the data in col.3 have been collected: symbols 
stand for H$=$H.E.S.S., M$=$MAGIC, W$=$Whipple.

\noindent
{\it Col.6:} corrected 0.2-2 TeV photon spectral index.

\noindent
{\it Col.7:} corrected $>$0.2 TeV flux.

\noindent
{\it Col.8:} 5 GHz flux density (in Jy): data are from the NED, except for 1ES~0347-121 (Fossati et al. 1998) 
and H~2356-309 (Costamante \& Ghisellini 2002).

\noindent
{\it Col.9:} 2-10 keV flux (in erg cm$^{-2}$ s$^{-1}$): the data come from the papers listed in {\it Col.11} 
or from references quoted therein; when necessary, fluxes have been converted from {\it R}XTE/ASM count rates 
using the conversion 1 ct/s $=$3.33E$-$10 erg cm$^{-2}$ s$^{-1}$. 

\noindent
{\it Col.10:} minimum 2-10 keV flux (in erg cm$^{-2}$ s$^{-1}$).

\noindent
{\it Col.11:} Reference for VHE $\gamma$-ray data, and for the baseline X-ray flux when the latter was not 
taken from the NED -- 
Al+07a: Albert et al. 2007a;
Al+07b: Albert et al. 2007b;
Al+07c: Albert et al. 2007c;
Al+06a: Albert et al. 2006a;
Al+06b: Albert et al. 2006b;
Al+07d: Albert et al. 2007d;
Ah+05a: Aharonian et al. 2005a; P+99: Perlman et al. 1999; 
Ah+08:  Aharonian et al. 2008; D+01: Donato et al. 2001; 
Ah+05b: Aharonian et al. 2005b;
Hor+02: Horan et al. 2002; S+97: Sambruna et al. 1997; 
Ah+07c: Aharonian et al. 2007c; P+96: Perlman et al. 1996; D+05: Donato et al. 2005;
Ah+06a: Aharonian et al. 2006a; W+84: Wood et al. 1984; 
Al+06c: Albert et al. 2006c;
Ah+07a: Aharonian et al. 2007a;
Ah+07b: Aharonian et al. 2007b;
Al+07e: Albert et al. 2007e;
Al+07f: Albert et al. 2007f.


\end{table*}

\section{Intergalactic absorption}

The cross section for the reaction $\gamma \gamma$$\rightarrow$$e^{\pm}$  is  
(Heitler 1960), 
\begin{eqnarray}
\lefteqn{
\sigma_{\gamma \gamma}(E, \epsilon) ~=~ 1.25 \times 10^{-25} ~(1-\beta^2)~~\times }
	\nonumber\\
 & & ~~~~~~~~~~~~~~~
\times ~~ 2\,\beta\,(\beta^2-2) ~+~ (3-\beta^4) ~{\rm ln} {1+\beta \over 1-\beta} \, , 
\end{eqnarray}
where $\beta$$\equiv$$\sqrt{1-(m_ec^2)^2/E\epsilon}$. By calling $n(\epsilon)$
the number per unit volume of background photons with  energy equal to
$\epsilon$ at the current position, the corresponding optical depth 
due to the attenuation between the source redshift, $z_e$, and the Earth, is 
\begin{eqnarray}
\lefteqn{
\tau_{\gamma \gamma}(E) ~=~ {c \over H_0} \int_0^{z_e} \sqrt{1+z}~ {\rm d}z ~ 
\int_0^2 {x \over 2} {\rm d}x ~~~ \times} 
                \nonumber\\
 & & ~~~~~~~~~~~~~~~
\times ~ \int_{2(m_ec^2)^2 \over Ex(1+z)^2}^\infty n(\epsilon)~~ \sigma_{\gamma \gamma}\bigl(2xE \epsilon 
(1+z)^2\bigr) ~~ 
{\rm d}\epsilon\,,
\end{eqnarray}
where $x$$\equiv$$(1$$-$${\rm cos}\,\theta)$, $\theta$ being the angle between the photons, and $H_0$ 
is the Hubble constant. Eq.(1) assumes $\Omega_0$$=$1 and no redshift evolution of $n(\epsilon)$ -- the 
latter assumption being adequate within the relatively low redshifts relevant to this paper. For 
demonstration purposes let us assume, following Stecker et al. (1992), that the local IBL has a power-law 
form, $n(\epsilon)$$\propto$$\epsilon^{-2.55}$: then the above integral yields $\tau(E,z)$$\propto$$E^
{1.55}$$z_s^{\eta}$ with $\eta$$\sim$1.5 (see Stecker et al. 1992). This calculation shows an important 
property of $\tau_{\gamma\gamma}$: it depends both on the distance traveled by the hard photon (hence on $z$) 
and on its energy.
The expected VHE $\gamma$-ray flux at Earth will be, of course: $F(E)$$=($d$I/$d$E)\,e^
{-\tau_{\gamma \gamma}(E)}$ (differential) and $F($$>$$E)$$=$$\int_E^\infty ($d$I/$d$E^\prime)\,e^{-
\tau_{\gamma \gamma}(E^{\prime})}dE^\prime$ (integral).

Along the same path, but using recent data on galaxy luminosity functions and redshift evolution, 
Stecker et al. (2006, 2007) made a detailed evaluation of the IBL density as a function of both energy 
and redshift for 0.003 eV $\leq$ $\epsilon$ $\leq$13.6 eV and 0$<$$z$$<$6 in the $\Lambda$CDM universe with 
$\Omega_\Lambda$$=$0.7 and $\Omega_m$$=$0.3. Using their calculated IBL photon densities, they 
calculated $\tau_{\gamma \gamma}$ for $\gamma$-rays with energies 0.004$\leq$($E$/TeV)$\leq$100 emitted 
by sources at redshift 0$<$$z$$<$5. 

Stecker \& Scully (2006) fitted such $\tau_{\gamma \gamma}(E,z)$ to 
a form that was assumed to be logarithmic in $E$ in the energy range 0.2 TeV$\leq$$E$$\leq$2 TeV and linear 
in $z$ for 0.05$\leq$$z$$<$0.4, i.e., $\tau_{\gamma\gamma}(E,z)$$=$$(A+Bz)$$+$$(C+Dz)\,{\rm ln}(E/{\rm 
TeV})$. Consequently, if the intrinsic source spectrum can be described as a power law, $F_s(E)$$=$$KE^
{-\alpha}$, in the band between 0.2 TeV and 2 TeV, the observed emission can
be approximated as $F_o(E)$$=$$Ke^{-(A+Bz)}$$E^{-(\alpha+
C+Dz)}$, i.e.,  the spectrum will be attenuated by a factor $e^{-(A+Bz)}$ and steepened by 
$\Delta \alpha$$=$$C+Dz$. The numerical values given by Stecker \& Scully (2006) are A= $-$0.346 ($-$0.475), 
B=16.3 (21.6), C= $-$0.0675 ($-$0.0972), and D=7.99 (10.6) in the Stecker et al. (2006) baseline (fast) 
evolution model. 

\begin{figure*}
\vspace{7.6cm}

\includegraphics{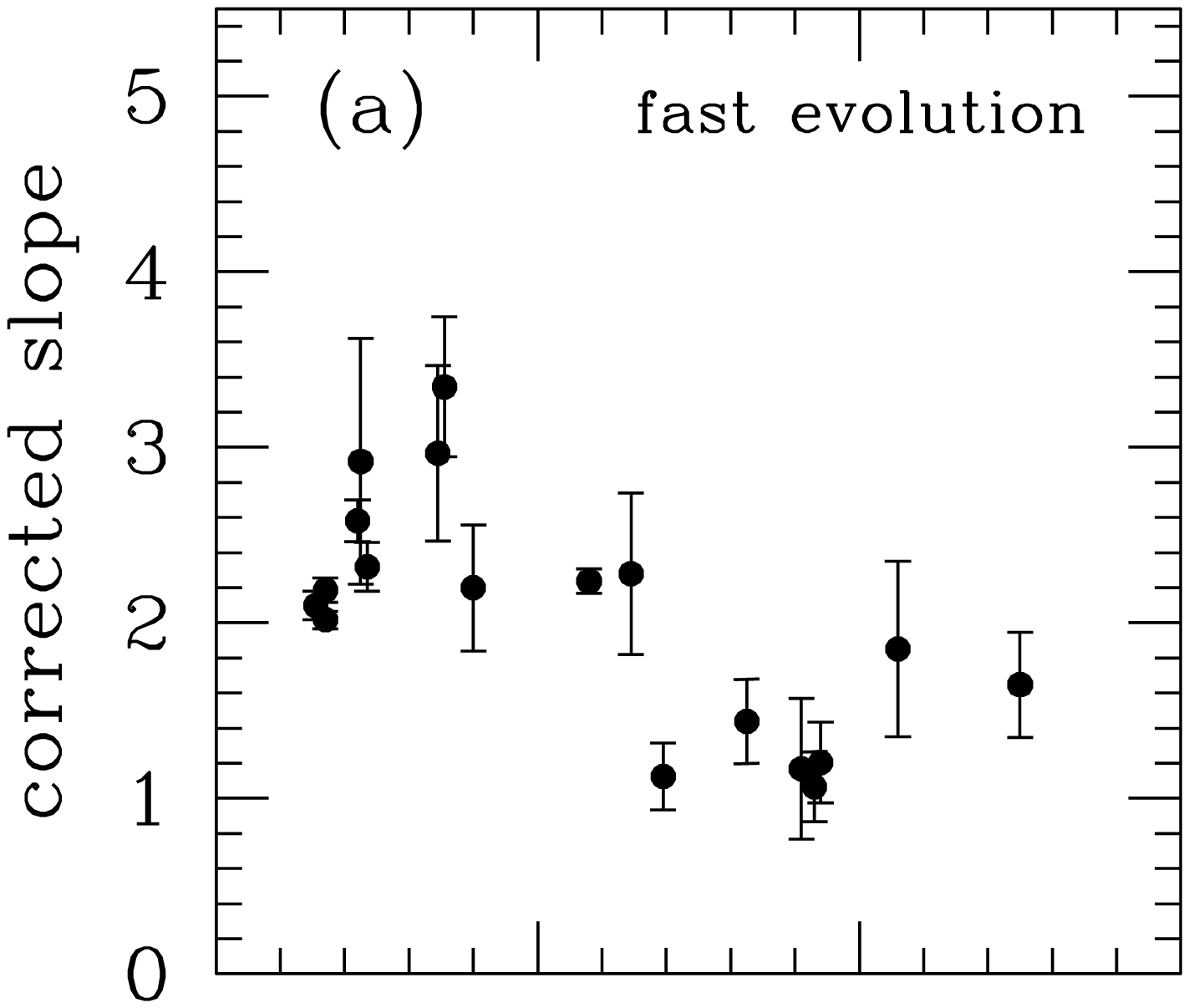}
\includegraphics{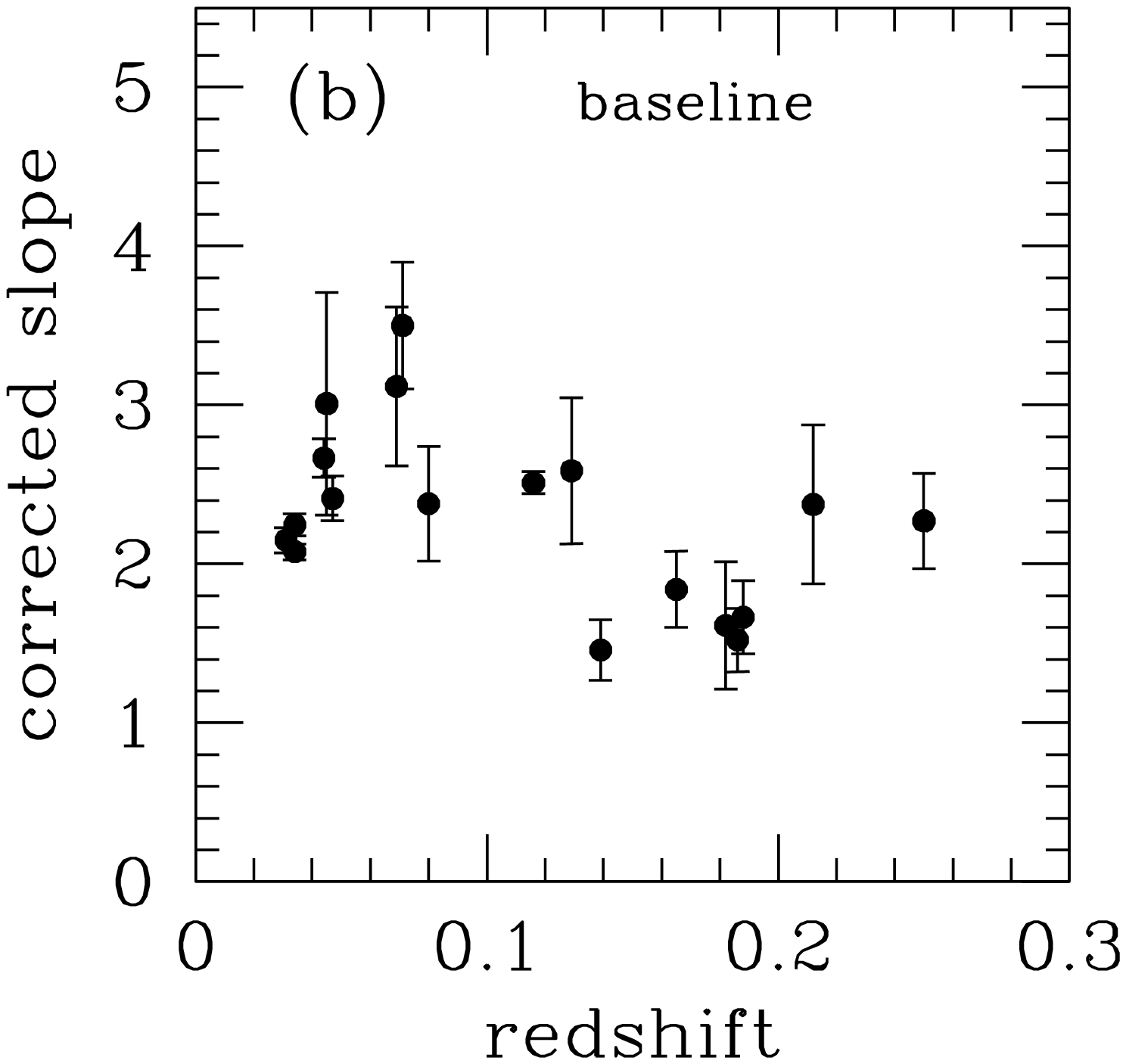}
\includegraphics{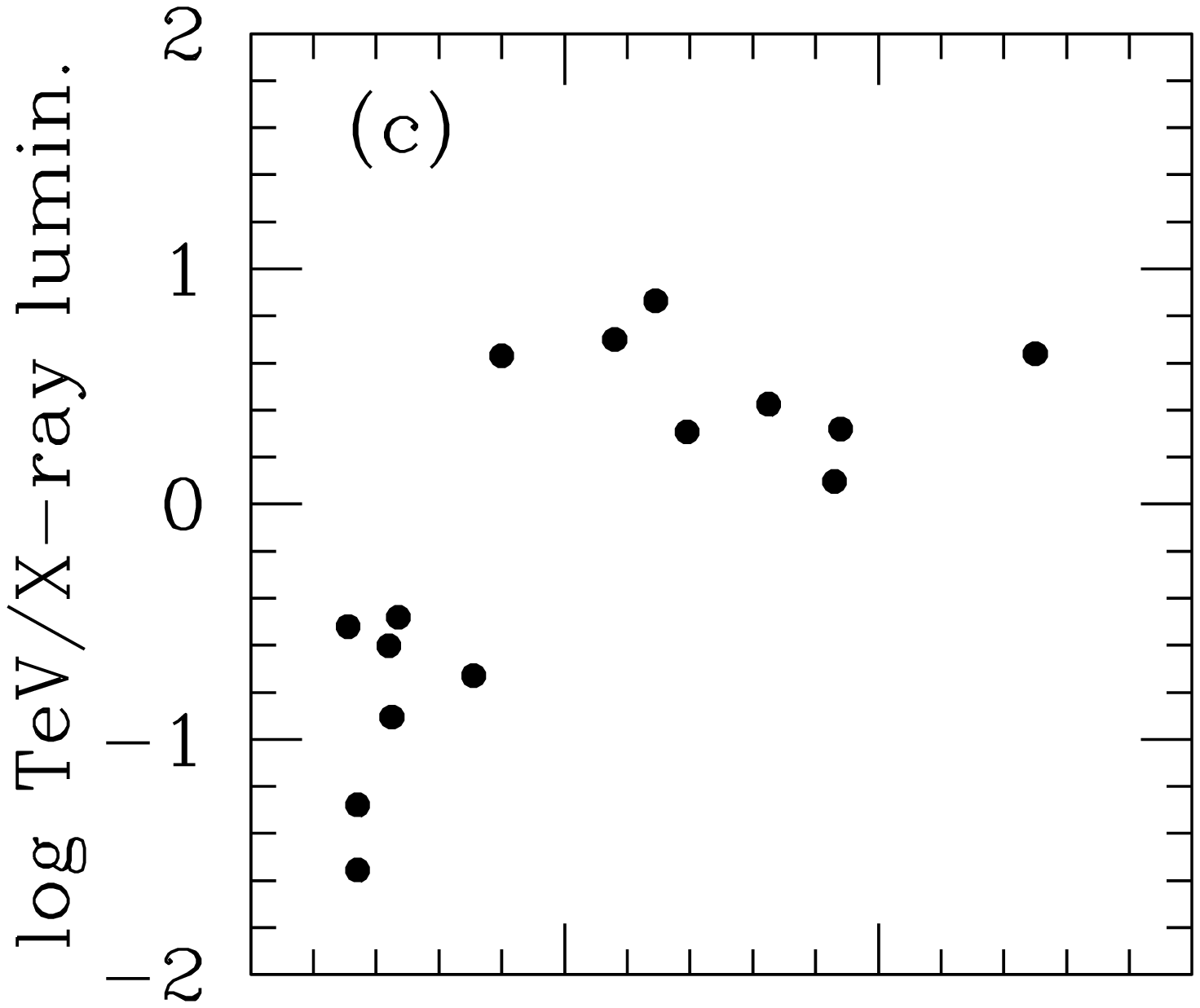}
\includegraphics{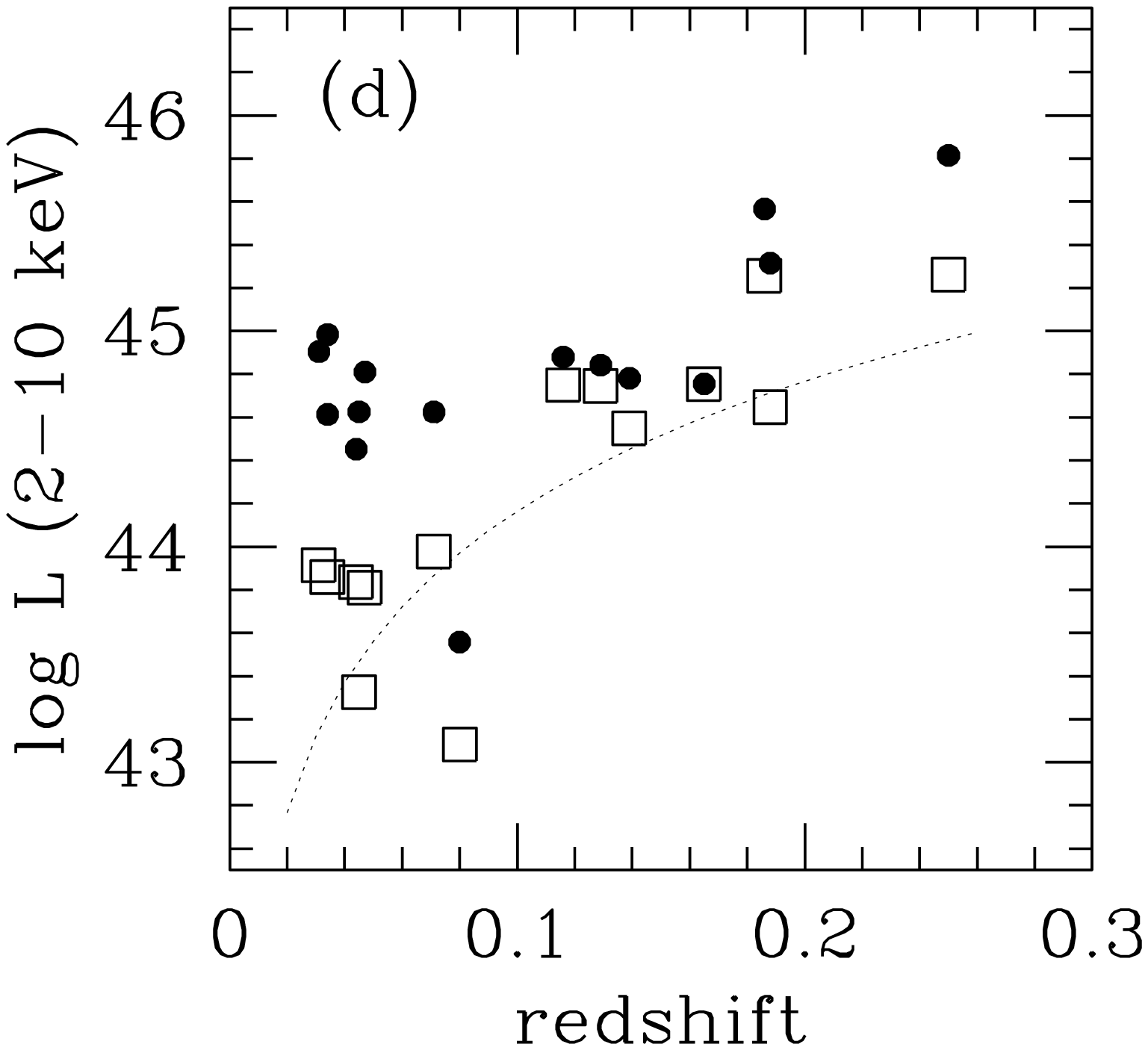}
\includegraphics{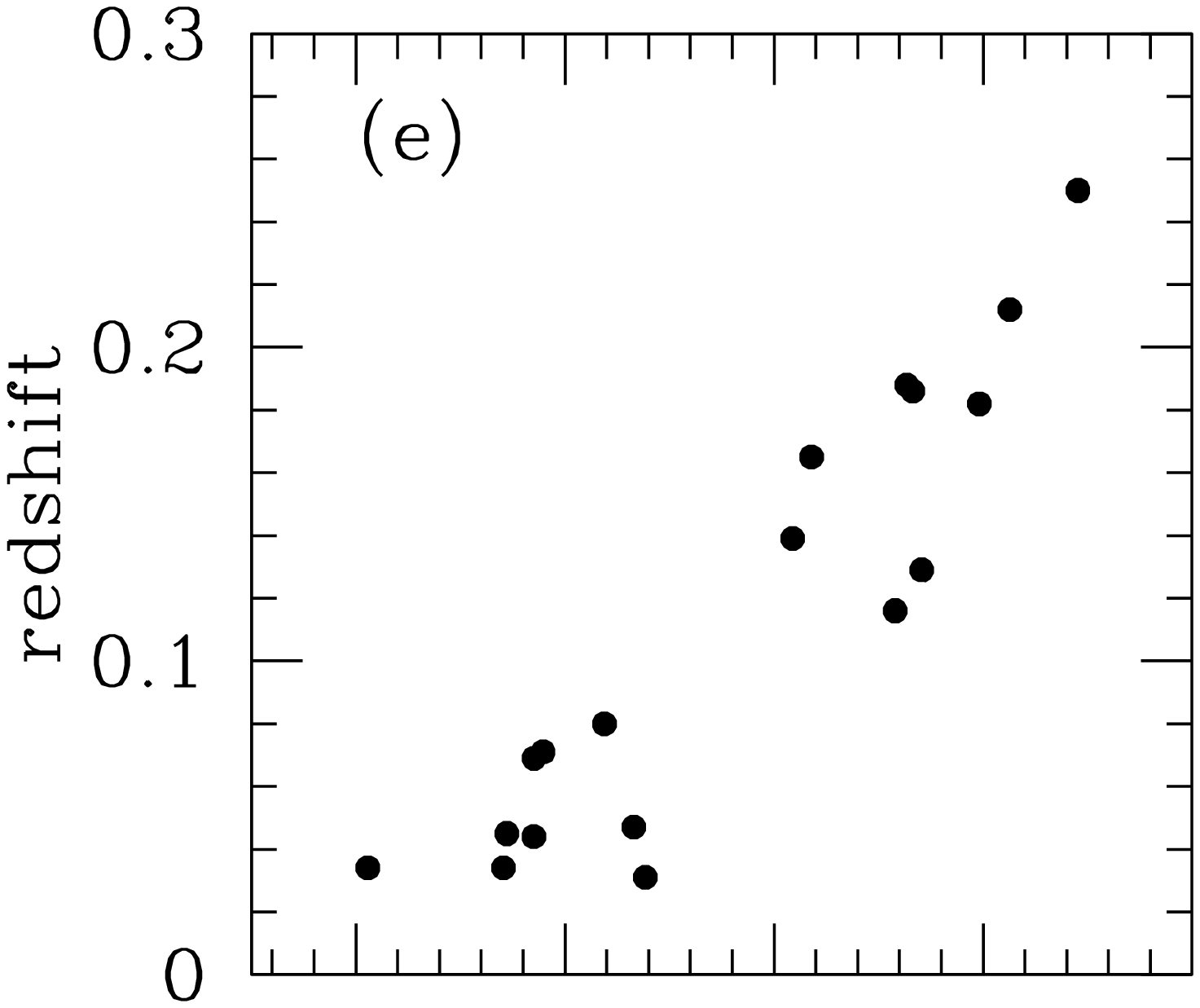}
\includegraphics{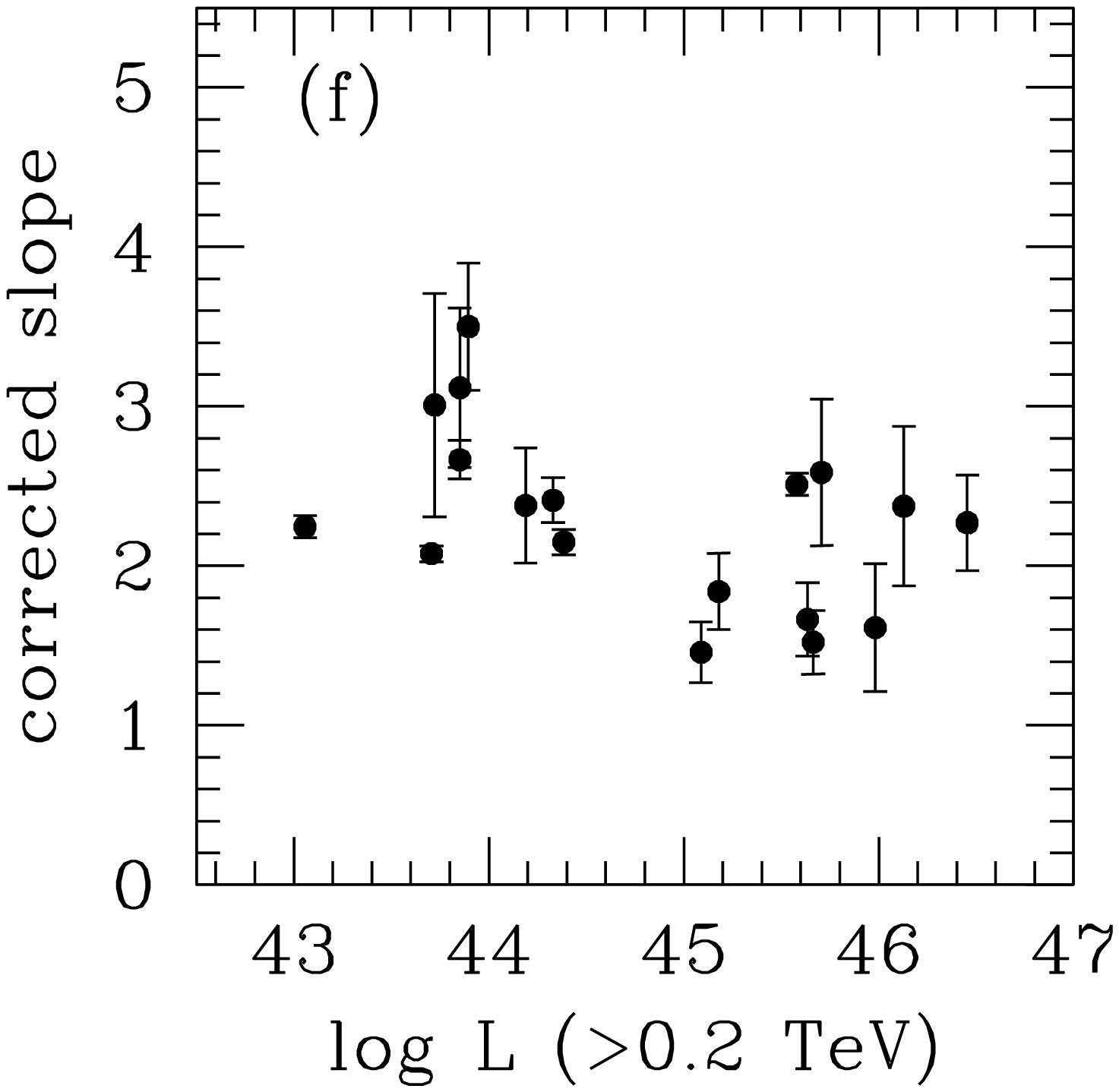}
\includegraphics{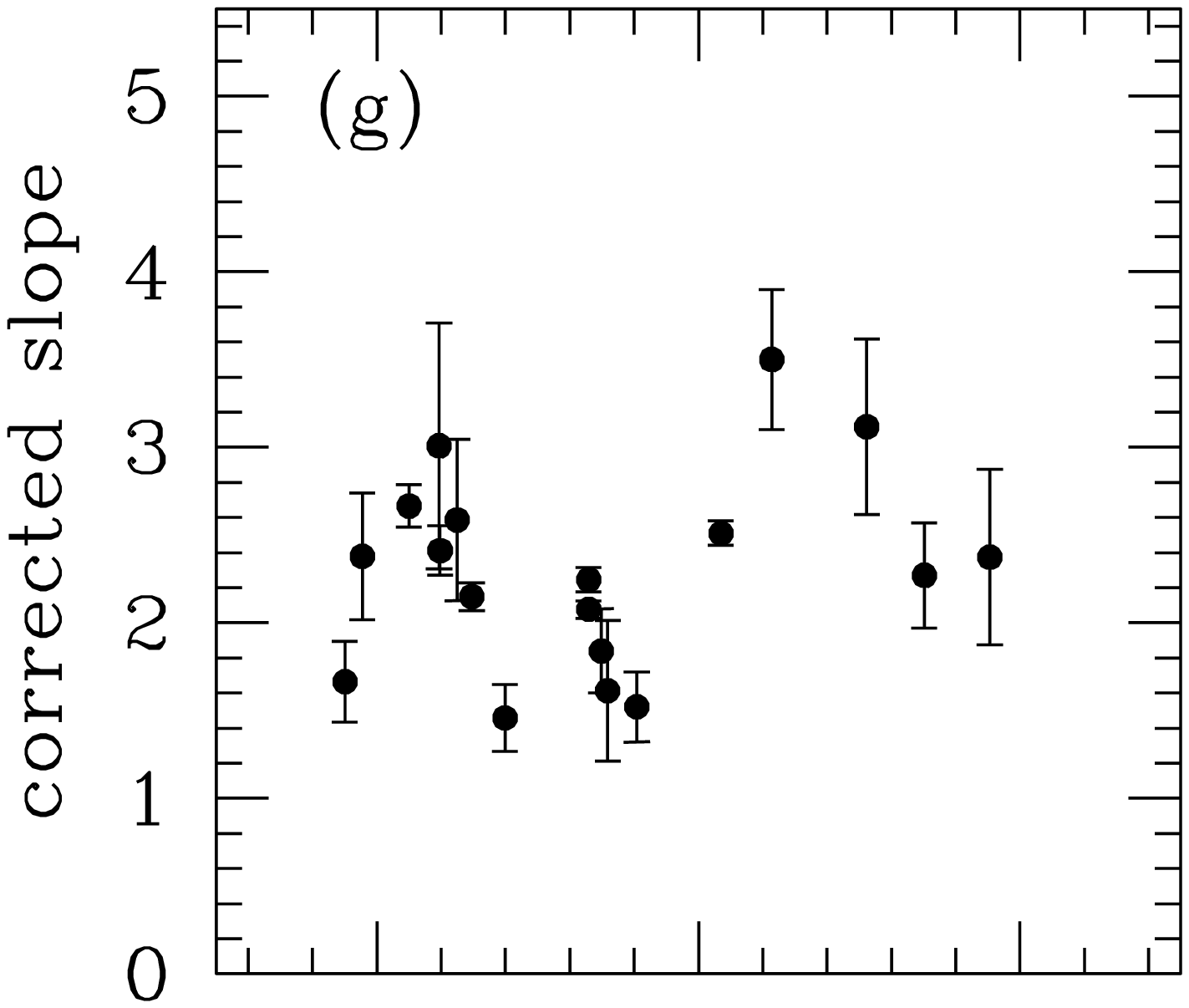}
\includegraphics{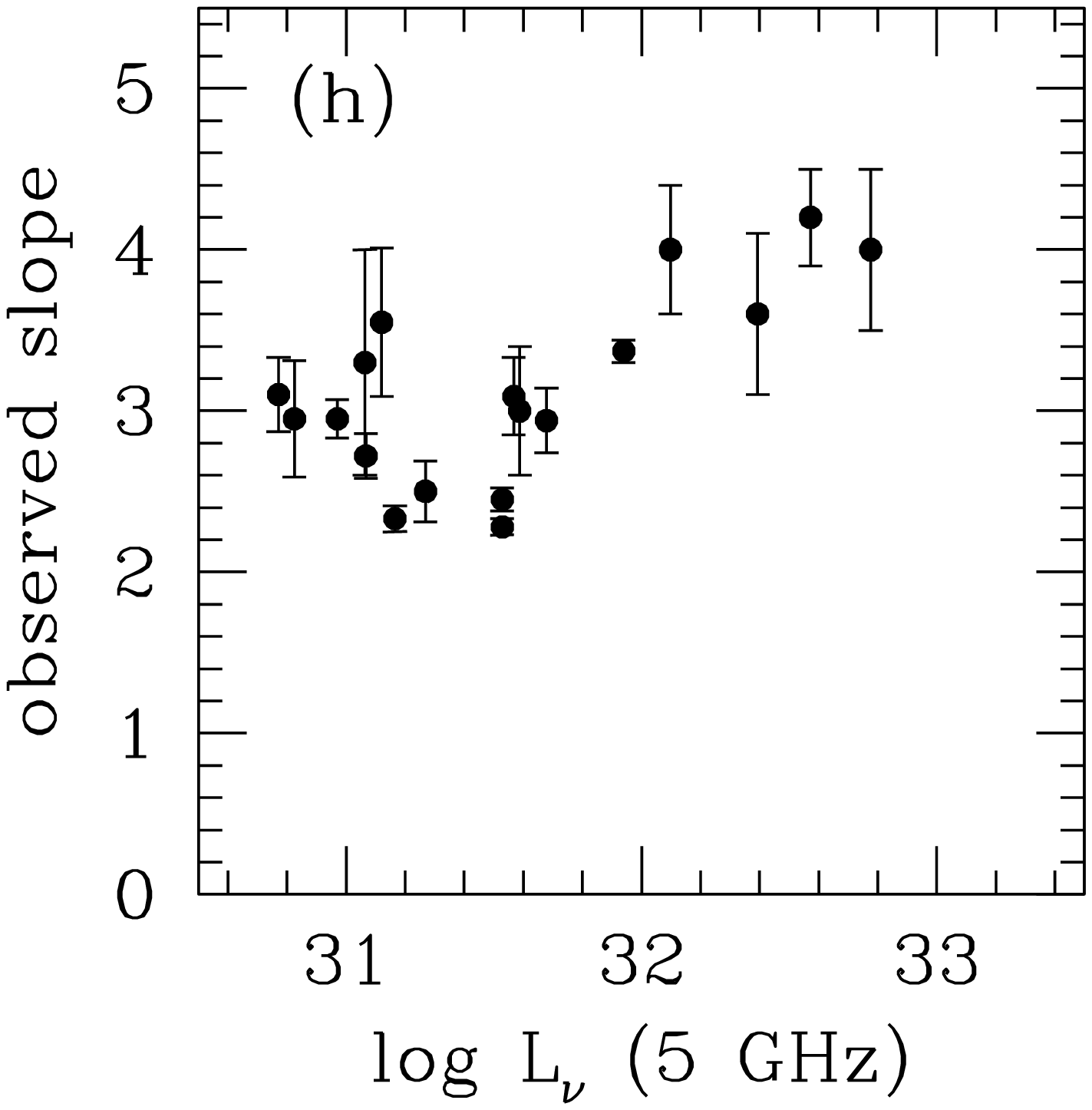}

\caption{ Parameter trends for the TeV blazars in Table 1. 
{\it (i)} IBL-corrected VHE slope vs redshift. The spectral slopes have been corrected for IBL 
absorption according to the `fast evolution' {\it (a)} and `baseline' {\it (b)} IBL models of 
Stecker et al. (2006) as parameterized by Stecker \& Scully (2006). 
{\it (ii)} VHE to 2-10 keV luminosity ratio versus redshift {\it (c)}, and 2-10 keV luminosity 
versus redshift {\it (d)}. Given the modest redshifts encompassed by the current sample of TeV 
blazars, distances have been computed according to $D(z)$$=$$cz/H_0$ (with H$_0$$=$72 km s$^{-1}$ 
Mpc$^{-1}$), and luminosities have not been k-corrected. The VHE luminosities are 
IBL-corrected using the baseline model of Stecker \& Scully (2006). In panel {\it (d)}, the filled 
dots represent X-ray data taken simultaneously with the TeV measurements, whereas the empty squares 
represent historical low flux states; the thin dotted curve represents the 'visibility function' 
for a 2-10 keV instrumental sensitivity of 7$\times$10$^{-12}$ erg cm$^{-2}$ s$^{-1}$ ($\sim$0.3 
mCrab). 
{\it (iii)} TeV luminosity versus redshift {\it (e)} and versus TeV spectral slope {\it (f)}.
Luminosities and slopes are IBL-corrected using the baseline model of Stecker \& Scully (2006).  
{\it (iv)} Intrinsic {\it (g)} and observed {\it (h)} TeV slopes versus 5 GHz monochromatic 
radio luminosities.
}
\end{figure*}

In Fig.2a,b we plot the IBL-corrected slopes versus redshift, using the baseline (Fig.2a) and fast 
(Fig.2b) evolution models of Stecker \& Scully (2006) 
	\footnote{Mrk 421, Mrk 501, and PKS 2344+514 are all at $z$$<$0.05, 
		    which is the lower limit for the analytic approximation given in Stecker 
		    \& Scully (2006): this, although it means slightly overestimating the 
		    (modest) true absorption for these sources, will in no way affect the 
		    main conclusions of this paper.}. 
A loose trend of {\it flatter} slopes at higher redshift is apparent now: from Table 1 we derive $<
$$\alpha_\gamma^{\rm corr}$$>$$=$2.62$\pm$0.16 for $z$$<$0.1 and $<$$\alpha_\gamma^{\rm corr}$$>$$=
$1.97$\pm$0.15 for $z$$>$0.1 (using the baseline correction; the fast-evolution correction only 
enhances the effect). In this paper we suggest that the trend of intrinsic slope with redshift is 
induced by a luminosity effect -- the latter, in turn, resulting from a selection effect. A 
possible correlation of spectral flatness and luminosity with redshift was also pointed out by 
Stecker et al. (2007).

\section{A luminosity effect}

Let us consider which emission conditions in blazars can lead to flatter TeV slopes. The standard 
synchrotron-Compton emission model of blazars (e.g., Jones et al. 1974; Maraschi et al. 1992; 
Mastikiadis \& Kirk 1997) gives us a clue on the link between active phase and TeV spectral shape in 
blazars. Its basic feature is that, if the distribution function of the emitting electrons changes 
at the highest energies, correlated flaring at X-ray and TeV energies are produced, with the highest 
energy electrons producing X-rays via snchrotron and the TeV radiation via IC scattering. 

A well-known example of this correlation is provided by Mrk 501 (Pian et al. 1998): during the giant 
flare of April 1997, its (apparent) bolometric luminosity increased by a factor of $\sim$20, its 
synchrotron and IC peaks shifted to higher energies by, respectively, a factor $\magcir$100 (into the 
hard X-ray range, $\magcir$50--100 keV) and a factor $\magcir$10 (Klein-Nishina limited, into 
the sub-TeV range), and $f_{2-10\, {\rm keV}}$ increased by a factor of $\sim$3. As the shift of the 
synchrotron peak by $\sim$2 orders of magnitude could not be due to variations of the magnetic field 
and/or the Doppler factor because enormous variations would have been demanded on either quantity that 
would have revealed themselves by affecting other parts of the spectrum (which was not observed), the 
luminosity and spectral variation could be explained by invoking an injection, continuous throughout 
the duration of the burst, of energetic electrons that became superposed to the baseline distribution 
of emitting electrons responsible for the quiescent emission of Mrk 501. The former had a spectrum that 
was flatter than the latter, and extended to higher (by a factor $\sim$10--30) energies (see Pian et al. 
1998). The resulting spectral energy distribution (SED) involved, w.r.t. the baseline case, a 
higher power output, a higher Compton/synchrotron power ratio, and both the synchrotron and Compton 
peaks located at higher energies: that meant a harder TeV slope and enhanced 2-10 keV and TeV fluxes. 
The hardening and increase of the TeV emission, simultaneously with an increase of the 2-10 keV emission, 
has been observed also in other blazars, down to very short (e.g., hours and minutes) time scales (e.g., 
see Mrk421 and Mrk501 in Table 1 and references cited therein). 

Because all the known TeV blazars, including Mrk 501, are of the HBL type (with one LBL exception, BL 
Lacertae), whose SEDs are well reproduced by the SSC model (Ghisellini et al. 1998), we expect that the 
basic features of Pian et al.'s (1998) model of the 1997 burst of Mrk 501 apply to TeV blazar in general. 
Hence, we do expect a correlation between harder TeV spectral shapes and more active emission states in 
(HBL) blazars. We then have to examine whether more distant sources are found, on the average, in more 
active states.

\section{A distance bias for active states?}

Within the theoretical SSC framework outlined above, two indicators of enhanced emission activity are 
{\it (i)} the TeV/X-ray luminosity ratio, and {\it (ii)} the X-ray enhancement. The former quantity, 
based on simultaneous data, represents the Compton/synchrotron power ratio. The latter quantity is 
defined as the ratio of the 2-10 keV power, measured simultaneously with the TeV observations, to its 
lowest historical level: it measures the increase of the synchrotron power, i.e. the strength of the 
activity. 
\smallskip 

\noindent
{\it (i)} In Fig.2c we plot the ratio of intrinsic TeV to X-ray luminosity versus redshift. In both bands, 
the luminosities are computed from the observed fluxes and redshifts (k-corrections do not significantly 
alter the results
	\footnote{  
                    The full-band k-correction, to be applied to the 2-10 keV luminosities and 
		    the 0.2-2 TeV luminosities, amounts to a factor $(1+z)^{\Gamma-2}$ if in 
		    the relavant band the spectrum is power law with intrinsic slope $\Gamma$.
		    For the blazars in Table 1, it is either $z$$<$0.1 or $\alpha_\gamma^{\rm corr}$$\sim$2 
		    (and typically $\alpha_{2-10\,{\rm keV}}$$\sim$2.2, see Donato et al. 2005), so the 
		    corresponding k-corrections to the X-ray and VHE $\gamma$-ray luminosities are small. 
                 }
), whereas 
the TeV luminosities are IBL-corrected using the baseline model of Stecker \& Scully (2006). Even though 
the fields of view (FoVs) of the different X-ray and Cherenkov detectors are different, since the measured 
blazar fluxes do originate from (effectively) point-like sources and no other (known) X-ray or TeV sources 
lie in the detectors' FoVs, in Fig.2c there is no need to account for the different FoVs. Furthermore, 
in the SSC framework the synchrotron and Compton emissions are essentially cospatial and hence are affected 
by the same beaming-related boost factor. Indeed there is a sharp correlation, suggesting an increasing 
importance of the Compton hump for more distant sources.
\smallskip

\noindent
{\it (ii)} In Fig.2d we plot the X-ray luminosities of our TeV blazars versus redshift (filled dots: 
fluxes measured simultaneously with the TeV measurements; open squares: faintest historical fluxes). 
The lower limit to the allowed $L_{2-10\,{\rm keV}}(z)$ distribution in Fig.2d is dictated by a 
combination of X-ray telescope sensitivities, exposure times, and source spectral shape: for sake of 
illustration, in Fig.2d we plot an effective limiting sensitivity of $f_{\rm s}$$=$0.7$\times$10$^{-11}$ 
erg cm$^{-2}$ s$^{-1}$. A trend is apparent in the data. Whereas the ground-state emissions (open 
squares) tend to be found close to the limiting curve, the TeV-selected X-ray emissions tend to lie 
significantly above it. The strongest enhancements, as defined here, occur for the nearest sources 
($z$$<$0.1): this however stems from a selection effect, as fainter fluxes can be detected from more 
nearby, but otherwise similar, sources. For more substantial redshifts ($z$$>$0.1), we notice that 
the X-ray emissions tend to be more enhanced w.r.t. their ground states with increasing $z$: although 
noisy, there appears to be a link between such enhancements and redshift. 

Thus, both our activity indicators suggest that TeV sources further away tend to be observed during more 
active phases. Why? Our tentative answer -- a selection effect -- is based on the following argument. 
Moving to higher redshifts, X-ray and TeV measurements are both affected by geometrical flux dilution 
($\propto z^{-2}$), but the latter are also affected by IBL aborption ($\propto e^{-\tau_{\gamma 
\gamma}(z)}$). This implies that, in order for sources to be detected at both keV and TeV energies with 
a given pair of X-ray and Cherenkov detectors, the ratio of {\it intrinsic} TeV to 2-10 keV luminosities 
must increase with redshift starting from a threshold redshift that depends on the sensitivity of the 
two instruments. Recalling the behavior of the SSC model for Mrk 501's giant 1997 flare by Pian et al. 
(1998), this means observing sources that are in increasingly higher states of emission. 

This argument entails a testable prediction. If distant TeV blazars are observed during strong flares, 
then, because of the high electron energies involved, the electrons' scattering cross-section is reduced by 
the Klein-Nishina effect, which will cause the Compton peak to shift to higher energies more slowly 
than the synchrotron peak (which depends on the square of the electron energy). 
We then expect the TeV spectral shape to change quite little at the highest redshifts ($z$$>$0.1) -- 
or equivalently, through the $L_{\geq 0.2\,{\rm TeV}}(z)$ relation (see Fig.2e), at the highest 
luminosities ($L_{\geq 0.2\,{\rm TeV}}$$>$10$^{45}$ erg cm$^{-2}$ s$^{-1}$). Conversely, for less luminous 
sources (i.e., sources caught in less active states) the energetics suggests a still largely Compton regime 
of the upscattered photons, which allows the Compton peak to shift to higher energies at the same rate as 
the synchrotron peak: hence a spectral flattening is expected with increasing luminosity for the more 
nearby sources. The data seem to conform to this prediction (see Fig.2f), with the weaker sources showing 
a steeper intrinsic value ($\alpha_\gamma^{\rm corr}$$\sim$3) and the more powerful ones being spectrally 
revealed in the vicinity of the Compton peak ($\alpha_\gamma^{\rm corr}$$\sim$2).

The radio emission, on the other hand, is largely decoupled from the X-ray and TeV emissions. In the 
'canonical' model of blazar emission the high-energy (X-ray and $\gamma$-ray) emission, though constituting 
most of the output, is produced in a region of the blazar's jet that has to be compact in order to account 
for the short-time variability shown by the blazars. This makes the radio emission from such region strongly 
self-absorbed, so that the observed radio emission must originate in a much larger volume. Therefore the 
link between the radio emission and the X-/$\gamma$-ray region may be much more subtle and less direct than 
suggested by a simple application of the SSC model (e.g., Costamante \& Ghisellini 2002). As a consequence, 
the above arguments linking TeV slope and X-/$\gamma$-ray luminosity do not apply in the radio case: indeed, 
there is no correlation between intrinsic TeV slopes and radio luminosities (Fig.2g). There is, however, an 
overall trend of steeper {\it observed} slopes with increasing luminosities (Fig.2h), a consequence of more 
luminous objects being typically more distant and hence more IBL-affected at TeV energies, coupled with the 
essentially flat distribution of $\alpha_{\gamma}^{\rm corr}(L_{5\, {\rm GHz}})$.

\section {Discussion and Conclusion}

Within the quite modest statistics and redshift range spanned by known TeV blazars, we find -- contrary to 
some early expectations -- no obvious dependence of the observed TeV slopes on redshift. However we find 
that, once the TeV slopes and luminosities are corrected for intergalactic absorption, flatter slopes tend to 
be found in more distant, more luminous blazars. (A correlation of flatness and luminosity with redshift has 
been suggested also by Stecker et al. 2007.) We argue that this latter correlation descends from a selection 
effect: use of X-ray data, both simultaneous with TeV data and archival (when available), suggest that, for 
$z$$\magcir$0.1, more distant TeV blazars are in more active states (at the epoch of their observation); and 
more active states mean -- given the link between blazar activity and spectral shape observed in nearby objects 
(e.g., Mrk 501) -- harder TeV spectra. This activity--distance link, in turn, results from the selective effect 
of IBL absorption (that affects TeV, not keV, photons), which forces more distant objects to have higher intrinsic 
TeV/X-ray ratios in order to be detectable. In sum, the IBL appears to affect the blazars' TeV slope versus redshift 
distribution in two ways: first, by causing only objects in progressively more active states, hence with 
flatter intrinsic slopes, to be detected at higher redshifts; and then, by making the observed slopes steeper 
to the observer, more so for higher redshifts. The result of this twofold action of the IBL is -- within the 
relatively small redshift range probed, $z$$\mincir$0.25 -- a substantial lack of correlation between measured 
TeV slopes and redshift. 

It is clear, however, that if higher-luminosity/activity sources are detected at higher redshifts, the 
underlying Klein-Nishima regime will imply a substantially flat slope, $\alpha_\gamma^{\rm corr}$$\sim$2, over 
a large luminosity range -- hence over a relatively large redshift interval. Due to the redshift dependence 
of the IBL opacity, this entails a sharp correlation between observed spectral slope and redshift. We then 
predict that IACT observations of $z$$\magcir$0.2 blazars will observe progressively higher $\alpha_\gamma$'s 
with increasing redshift.

Of course, the above considerations are only as good as the adopted IBL model (Stecker et al. 2006) is. For a 
different IBL redshift dependence, the resulting distribution of points in Figs.2a,b would also be different. 
More exotic interpretations of the intergalactic extinction, e.g., in the framework of the mixing between 
photons and a very light axion-like particle (De Angelis et al. 2007), would predict more moderate flux dimming 
and spectral distortion than entailed by most IBL schemes (e.g., Kneiske 2004; Mazin \& Raue 2007). 
Reversing the arguments, if the maximum allowed hardness of TeV blazar spectra is known, the observed 
spectrum of a source with known distance in principle allows one to deduce an upper limit to the IBL. Aharonian et al. 
(2006b), 
using the H.E.S.S. observation of 1ES 1101-032 and arguing that $\alpha_\gamma^{\rm corr}$$\geq$1.5, place 
an upper limit to the IBL at 1.5$\mu$m that is close to its lower limit (i.e., the integration of galaxy 
counts), hence suggesting a 'low' IBL. Stecker et al. (2007) however, argue that $\alpha_\gamma^{\rm corr}$$
<$1.5 can be produced if particles in blazar jets are accelerated at relativistic shocks, and hence loosen 
Aharonian et al.'s (2006b) upper limit, so permitting a higher IBL density field. Finally, if both the intrinsic 
blazar spectrum and the IBL are assumed, the observed TeV spectrum can be used to estimate the distance 
to the source (e.g., Mazin \& Goebel 2007).

It will be important to increase the statics in order to clarify the situation. This can be achieved both by 
probing larger redshifts with deeper Cherenkov observations and by measuring TeV fluxes and spectra of nearby 
(z$<$0.1) blazars in different states of activity. The latter point, in particular, will enhance the statistics 
of the low-$z$ portion of Figs.1,2a,2b,2f by adding crucial information on the quiet states of (nearby) blazars.
As for the former point, the newly operating VERITAS telescope and the upcoming enhanced MAGIC and H.E.S.S. 
telescopes should improve the sensitivity of ground-based observations considerably, so adding to 
the number of known $z$$>$0.2 TeV blazars with spectral information above $\sim$0.1 TeV. The impending GLAST 
satellite, too, will have an importat role in such expansion: its (largely IBL-unaffected) operational band, 
$\sim$0.01-100 GeV (with a sensitivity higher, by a factor of $\sim$50, than its predecessor EGRET), will 
sample the blazar SED around the Compton peak, so in principle allowing the solution of the SSC model 
and hence the knowledge of the intrinsic slope above $\sim$0.1 TeV. 

Finally, we emphasize the need for future TeV blazar discoveries to be immediately followed-up in X-rays, 
that will help to validate or invalidate the main argument made in this paper, i.e. that more flaring sources 
are detected at higher redshifts. This, if confirmed, would lend further support for the SSC process in blazar 
emission.

\medskip
\noindent
{\it Acknowledgements.} We thank Elena Pian, Daniel Mazin, and an anonymous referee for useful 
comments and suggestions. We acknowledge the MAGIC collaboration for providing a stimulating, 
friendly, and effective work environment. 
\medskip

\noindent
{\it Note added in proof.} Subsequent to the acceptance of this paper, newly 
published data for the HBL blazar RGB~J0152+017 was added. This is reflected 
in the figures and text (and Table 1). The results and conclusions of the paper 
are not affected in any major way by the inclusion of the new data. 
\bigskip

\def\ref{\par\noindent\hangindent 20pt}

\noindent
{\bf References}
\vglue 0.2truecm

\ref{\small Aharonian, F.A., Akhperjanian, A.G., Barres de Almeida, U., et al. 
            (H.E.S.S. collab.) 2008, A\&A, in press (arXiv: 0802.4021)} 
\ref{\small Aharonian, F.A., Akhperjanian, A.G., Bazer-Bachi, A.R., et al. 
            (H.E.S.S. collab.) 2007a, A\&A, 470, 475 } 
\ref{\small Aharonian, F.A., Akhperjanian, A.G., Barres de Almeida, U., et al. 
            (H.E.S.S. collab.) 2007b, A\&A, 473, L25 }
\ref{\small Aharonian, F.A., Akhperjanian, A.G., Barres de Almeida, U., et al. 
            (H.E.S.S. collab.) 2007c, A\&A, 475, L9 } 
\ref{\small Aharonian, F.A., Akhperjanian, A.G., Bazer-Bachi, A.R., et al. 
            (H.E.S.S. collab.) 2006a, A\&A, 455, 461} 
\ref{\small Aharonian, F.A., Akhperjanian, A.G., Bazer-Bachi, A.R., et al. 
            (H.E.S.S. collab.) 2006b, Nature, 440, 1018 } 
\ref{\small Aharonian, F.A., Akhperjanian, A.G., Bazer-Bachi, A.R., et al. 
            (H.E.S.S. collab.) 2005a, A\&A, 442, 895} 
\ref{\small Aharonian, F.A., Akhperjanian, A.G., Aye, K.-M., et al. 
            (H.E.S.S. collab.) 2005b, A\&A, 436, L17} 
\ref{\small Albert, J., Aliu, E., Anderhub, H., et al. (MAGIC collab.) 2007a, ApJ, 663, 125 }  
\ref{\small Albert, J., Aliu, E., Anderhub, H., et al. (MAGIC collab.) 2007b, ApJ, 669, 862 } 
\ref{\small Albert, J., Aliu, E., Anderhub, H., et al. (MAGIC collab.) 2007c, ApJ, 662, 892} 
\ref{\small Albert, J., Aliu, E., Anderhub, H., et al. (MAGIC collab.) 2007d, ApJ, 666, L17 } 
\ref{\small Albert, J., Aliu, E., Anderhub, H., et al. (MAGIC collab.) 2007e, ApJ, 667, L21 }
\ref{\small Albert, J., Aliu, E., Anderhub, H., et al. (MAGIC collab.) 2007f, ApJ, 654, L119}
\ref{\small Albert, J., Aliu, E., Anderhub, H., et al. (MAGIC collab.) 2006a, ApJ, 648, L105} 
\ref{\small Albert, J., Aliu, E., Anderhub, H., et al. (MAGIC collab.) 2006b, ApJ, 639, 761} 
\ref{\small Albert, J., Aliu, E., Anderhub, H., et al. (MAGIC collab.) 2006c, 642, L119} 
\ref{\small Bloom, S.D., \& Marscher, A.P. 1996, ApJ, 461, 657}
\ref{\small Costamante, L., \& Ghisellini, G. 2002, MNRAS, 384, 56}
\ref{\small De Angelis, A., Roncadelli, M., \& Mansutti, O. 2007, Phys Rev D, 76, 121301 }
\ref{\small Donato, D., Ghisellini, G., Tagliaferri, G., \& Fossati, G. 2001, A\&A, 375, 739}
\ref{\small Donato, D., Sambruna, R.M., \& Gliozzi, M. 2005, A\&A, 433, 1163 }
\ref{\small Fossati, G., Maraschi, L., Celotti, A., Comastri, A., 
            \& Ghisellini, G. 1998, MNRAS, 299, 433}
\ref{\small Ghisellini, G., Celotti, A., Fossati, G., Maraschi, A., 
		\& Comastri, A. 1998, MNRAS, 301, 45}
\ref{\small Hauser, M.G., \& Dwek, E. 2001, ARA\&A 39, 249}
\ref{\small Horan, D., Badran, H.N., Bond, I.H., et al. 2002, ApJ, 571, 753} 
\ref{\small Jones, T.W., O'Dell, S.L., \& Stein, W.A. 1974, ApJ, 188, 353} 
\ref{\small Kino, M., Takahara, F., \& Kusunose, M. 2002, ApJ, 564, 97}
\ref{\small Kneiske, T.M., Bretz, T., Mannheim, K., \& Hartmann, D.H. 2004, A\&A, 413, 807 }
\ref{\small Maraschi, L., \& Ghisellini, G., \& Celotti 1992, ApJ, 397, L5 }
\ref{\small Mastikiadis, A., \& Kirk, J.G. 1997, A\&A, 320, 19)
\ref{\small Mazin, D., \& Goebel, F. 2007, ApJ, 655, L13 }
\ref{\small Mazin, D., \& Raue, M. 2007, A\&A, 471, 439 }
\ref{\small Perlman, E.S., Stocke, J.T., Schachter, J.F., et al. 1996, ApJS, 104, 251}
\ref{\small Perlman, E.S., Madejski, G., Stocke, J.T., \& Rector, T.A. 1999, ApJ, 523, L11}
\ref{\small Pian, E., Vaccanti, G., Tagliaferri, G., et al. 1998, ApJ, 492, L17}
\ref{\small Sambruna, R.M., George, I.M., Madejski, G., et al., 1997, ApJ, 483, 774}
\ref{\small Stecker, F.W. 1971, Cosmic Gamma Rays (Baltimore: Mono Book Corp.), 201 }
\ref{\small Stecker, F.W., Baring, M.G., \& Summerlin, E.J. 2007, ApJ, 667, L29 }
\ref{\small Stecker, F.W., \& Scully, S.T. 2006, ApJ, 652, L9}
\ref{\small Stecker, F.W., de Jager, O.C., \& Salamon, M.H. 1992, ApJ, 390, L49}
\ref{\small Stecker, F.W., Malkan, M.A., \& Scully, S.T. 2006, ApJ, 648, 774}
\ref{\small Stecker, F.W., Malkan, M.A., \& Scully, S.T. 2007, ApJ, 658, 1392}
\ref{\small Wood, K.S., Meekins, J.F., Yentis, D.J., et al. 1984, ApJS, 56, 507}

\end{document}